# An Efficient Optimal Energy Flow Model for Integrated Energy Systems Based on Energy Circuit Modeling in the Frequency Domain

Binbin Chen[a], Wenchuan Wu[a,*], Qinglai Guo[a], Hongbin Sun[a,*]

[a] *Department of Electrical Engineering, Tsinghua University, 100084, Beijing, China*

**ABSTRACT**

With more energy networks being interconnected to form integrated energy systems (IESs), the optimal energy flow (OEF) problem has drawn increasing attention. Extant studies on OEF models mostly utilize the finite difference method (FDM) to address partial-differential-equation (PDE) constraints related to the dynamics in natural gas networks (NGNs) and district heating networks (DHNs). However, this time-domain approach suffers from a heavy computational burden with regard to achieving high finite-difference accuracy. In this paper, a novel OEF model that formulates NGN and DHN constraints in the frequency domain and corresponding model compaction techniques for efficient solving are contributed. First, an energy circuit method (ECM) that algebraizes the PDEs of NGNs and DHNs in the frequency domain is introduced. Then, an ECM-based OEF model is formulated, which contains fewer variables and constraints than an FDM-based OEF model and thereby yields better solving efficiency. Finally, variable space projection is employed to remove implicit variables, by which another constraint generation algorithm is enabled to remove redundant constraints. These two techniques further compact the OEF model and bring about a second improvement in solving efficiency. Numerical tests on actual systems indicate the final OEF model reduces variables and constraints by more than 95% and improves the solving efficiency by more than 10 times. In conclusion, the proposed OEF model and solving techniques well meet the optimization needs of large-scale IESs.

**Nomenclature**

*Abbreviations*

| | |
|---|---|
| CGA | Constraint generation algorithm |
| CHP | Combined heat and power |
| DHN | District heating network |
| ECM | Energy circuit method |
| EPN | Electric power network |
| FDDP | Frequency domain distributed parameter |
| FDLP | Frequency domain lumped parameter |
| FDM | Finite difference method |
| GPM | Generalized phasor method |
| IDFT | Inverse discrete Fourier transform |
| IES | Integrated energy system |
| NGN | Natural gas network |
| NGU | Natural gas unit |
| OEF | Optimal energy flow |
| PDE | Partial differential equation |
| TDDP | Time domain distributed parameter |
| TPU | Thermal power unit |
| VSP | Variable space projection |

*Variables*

| | |
|---|---|
| $x$ | Space |
| $t$ | Time |
| $U$ | Voltage |
| $I$ | Current |
| $p$ | Pressure |
| $m$ | Mass flow |
| $T$ | Temperature |
| $h$ | Heat flow/heat power |
| $v$ | Flow velocity |
| $\rho$ | Density |
| $P$ | Active power |

*Parameters*

| | |
|---|---|
| $R$ | Resistance |
| $L$ | Inductance |
| $C$ | Capacitance |
| $G$ | Conductance |
| $Y$ | Admittance |
| $Z$ | Impedance |
| $Z_c$ | Characteristic impendence |
| $K$ | Controlled source |
| $\omega$ | Angular frequency |
| $l$ | Length |
| $S$ | Cross section area |
| $D$ | Diameter |
| $\alpha$ | Inclined angle |
| $\lambda$ | Friction coefficient |

* Corresponding author. Present address: Dept. Electrical Engineering, Tsinghua University, 100084, Beijing, China. Tel.: +86 137 0107 3689; Fax: +86 10 6278 3086–800. E-mail address: shb@tsinghua.edu.cn.

| | | | |
|---|---|---|---|
| $\mu$ | Heat dissipation coefficient | dot above | Phasor/frequency variables |
| $\phi$ | Heat transmission factor | $X[i,:]$ | Retrieve the $i$-th row of matrix $X$ as a vector |
| $\gamma$ | Propagation constant | $X[:,j]$ | Retrieve the $j$-th column of matrix $X$ as a vector |
| $r$ | Energy coupling ratio | $X[i,j]$ | Retrieve the entry of matrix $X$ at the $i$-th row and $j$-th column |
| $\varepsilon$ | Penalty coefficient | | |
| $u$ | Cost coefficient | diag($\cdot$) | Convert a vector to a diagonal matrix |
| $c_s$ | Sonic speed | Re($\cdot$) | Retrieve the real part |
| $c_p$ | Specific heat capacity | Im($\cdot$) | Retrieve the imaginary part |
| $g$ | Gravitational acceleration | | |
| $N$ | Number | | |
| $\Omega$ | Set of indices | | |
| $\mathbf{A}$ | Node-branch incidence matrix | | |
| $\mathbf{A}_+$ | Node-outflow-branch incidence matrix | | |
| $\mathbf{A}_-$ | Node-inflow-branch incidence matrix | | |
| $\mathbf{PTDF}$ | Power transfer distribution factor matrix | | |
| $\mathcal{A},\mathcal{B},\mathcal{C},\mathcal{D}$ | Transmission parameters of a two-port network | | |

*Superscripts and subscripts*

| | |
|---|---|
| $\bar{\cdot}$ | Base value obtained from historical values |
| $\hat{\cdot}$ | Intermediate optimal solution during iterations |
| $\cdot'$ | Per-unit-length distributed parameter |
| $\cdot^{(\tau)}$ | $\tau$-th time point |
| $\cdot^{(\kappa)}$ | $\kappa$-th frequency component |
| $\cdot^{ub}$ | Upper bound |
| $\cdot^{lb}$ | Lower bound |
| $\cdot^{rampD}$ | Down ramping limit |
| $\cdot^{rampU}$ | Up ramping limit |
| $\cdot_e$ | EPN/electric circuit |
| $\cdot_g$ | NGN/hydraulic circuit |
| $\cdot_h$ | DHN/thermal circuit |
| $\cdot_b$ | Branch |
| $\cdot_n$ | Node/bus |
| $\cdot_l$ | Pipeline/transmission line |
| $\cdot_{bf}$ | Branch end at "from" side |
| $\cdot_{bt}$ | Branch end at "to" side |
| $\cdot_{tpu}$ | Thermal power unit |
| $\cdot_{ngu}$ | Natural gas unit |
| $\cdot_{chp}$ | Combined heat and power unit |
| $\cdot_{wt}$ | Wind turbine |
| $\cdot_{gb}$ | Gas boiler |
| $\cdot_{hp}$ | Heat pump |
| $\cdot_{gw}$ | Gas well |
| $\cdot_{ld}$ | Load |
| $\cdot_{ht}$ | Historical time point |
| $\cdot_{dt}$ | Dispatch time point |
| $\cdot_f$ | Frequency component |
| $\cdot_{vsc}$ | Violated security constraint |
| $\cdot_i, \cdot_j$ | Index |

*Other Notations*

| | |
|---|---|
| bold font | Matrix/vector |

## 1. Introduction

### 1.1 Background

As it is gradually recognized that integrated energy systems (IESs) are capable of unlocking potential flexibility by shifting across different energy sectors [1], the interconnection of various energy networks to operate as a whole is increasing. To sufficiently benefit from this joint operation, it is essential to build an optimal energy flow (OEF) model that formulates coordinated dispatch schedules for the different energy networks in an IES [2].

### 1.2 Literature review

Although various energy carriers are considered in a general IES, we focus on IESs that consist of only electric power networks (EPNs), natural gas networks (NGNs), and district heating networks (DHNs) in this work. Regarding the OEF modeling for such IESs, several related works have been published, which are summarized as follows.

As pioneering works on OEF modeling, an optimal power-gas flow model is formulated in [3] to eliminate gas supply shortages, and another optimal power-heat flow model is presented in [4]-[5] to reduce wind power curtailment. By incorporating these elements, a comprehensive OEF model considering power, gas, and heat is developed in [6]. The advantages brought by coordination across energy sectors are clarified in these works.

Based on these fundamental OEF models, more complex factors are considered in later research. Some studies focus on improving the mathematical properties of OEF models: [7] and [8] utilize piecewise linearization and an extended convex hull to deal with nonlinear terms in OEF models, respectively. Considering uncertainties in IESs, a stochastic OEF model, a robust OEF model, and an OEF model with chance constraints are studied in [9], [10], and [11], respectively. In [12] and [13], OEF models with integrated demand response are discussed to exploit more flexibility resources in IESs. To address the data privacy issue brought by multiple managing entities in an IES, [14] and [15] solve the OEF model in a decentralized manner by using the alternating direction method of multipliers and generalized Benders decomposition, respectively. In addition, [16] proposes a multi-objective OEF model that simultaneously optimizes the operating cost and carbon emissions.

In addition to the above works, another important branch of OEF modeling is considering the dynamic characteristics of NGNs and DHNs in the OEF model. The fundamental OEF models in [3]-[6] are all formulated with steady-state equations of NGNs and DHNs, which brings the following two issues. (1) Security issue: From one steady state to another, there is a long dynamic process in NGNs and DHNs, during which the transient states are not considered by the steady-state OEF model. Thus, a feasible steady-state OEF solution may not reflect real feasibility in actual operation [17]. (2) Economic issue: The steady-state OEF model requires a real-time balance between energy supply and demand, which is an overly tight constraint and restricts the flexibility utilization of NGN and DHN inertias. With proper control, a temporary and limited mismatch between gas/heat supply and demand only results in an acceptable fluctuation of pressure/temperature. This enables the dynamic characteristics of NGNs and DHNs to work as energy storage to provide additional flexibility [18]-[19].

Compared with using approximate methods such as the line pack model [20]-[21] to develop dynamic OEF models, a more general and accurate approach is algebraizing the partial differential equations (PDEs) that formulate the dynamic processes in NGNs [22] and DHNs [23] and adding them as equality constraints to the optimization problem. To this end, the finite difference method (FDM) that algebraizes PDEs in the time domain is broadly adopted. In [24]-[26], FDMs with different difference schemes, including the explicit Euler scheme, implicit Euler scheme, and Lax-Wendroff scheme, are employed to develop dynamic optimal power-gas flow models, and similar works are implemented to derive dynamic optimal power-heat flow models in [27] and [28].

However, high precision of the FDM is achieved at the cost of introducing numerous mesh points on both time and space dimensions [29]. As a consequence, FDM-based dynamic OEF models have significantly more variables and constraints than steady-state OEF models, which degrades the solving efficiency. To address this issue, some other algebraization methods for dynamic OEF models have been explored. In [30]-[31], a method that adopts the Laplace transform to convert the time-domain PDEs of NGNs and DHNs into symbolic algebraic equations containing Laplacian $s$ in the complex frequency domain is proposed. The corresponding dynamic optimal power-heat flow model is observed to have fewer variables and constraints [31]. Nevertheless, the intractable symbolic calculation involving the Laplacian limits its applications in complex IESs. In [29] and [32], a generalized phasor method (GPM) that uses phasor representation to model NGNs and DHNs with a series of numeric algebraic equations in the frequency domain is proposed. The GPM is valid for reducing the sizes of dynamic OEF models and tractable for calculation, but it involves some rough simplifications and fails to derive network models, which restricts its further improvement.

*1.3 Work and contributions*

Moving toward a more efficient dynamic OEF model, the work in this paper is implemented in three steps. First, by extending the circuit modeling of transmission lines in EPNs to pipelines in NGNs and DHNs, an energy circuit method (ECM) that formulates gas and heat dynamics by numeric algebraic equations in the frequency domain is introduced, which is inherited from our previous works published in Chinese [33]-[34]. By combining this circuit representation with topological relationships, dynamic network models of NGNs and DHNs in matrix form are derived. Then, an ECM-based OEF model is developed, which replaces time-domain PDE constraints with frequency-domain energy circuit constraints and introduces time-frequency conversion constraints to couple the variables in these two domains. Finally, variable space projection is employed to express the implicit variables in the ECM-based OEF model as linear combinations of the explicit variables, and then a constraint generation algorithm is enabled to recognize critical security constraints. The corresponding contributions are as follows.

(1) The proposed ECM-based OEF model contains fewer variables and constraints than the conventional FDM-based OEF model, which yields better solving efficiency. Since the ECM is an accurate method similar to the FDM with regard to algebraization, the quality of the obtained dispatch schedules is still guaranteed.

(2) A technique of variable space projection is proposed to remove implicit variables that occupy the majority of all variables in the ECM-based OEF model. This technique not only reduces the variable count in the model but also enables another constraint generation algorithm.

(3) Based on the OEF model after variable space projection, a technique of constraint generation is proposed to remove redundant security constraints that can be overridden by other critical ones. This technique further compacts the model in the constraint dimension and realizes a second improvement in solving efficiency.

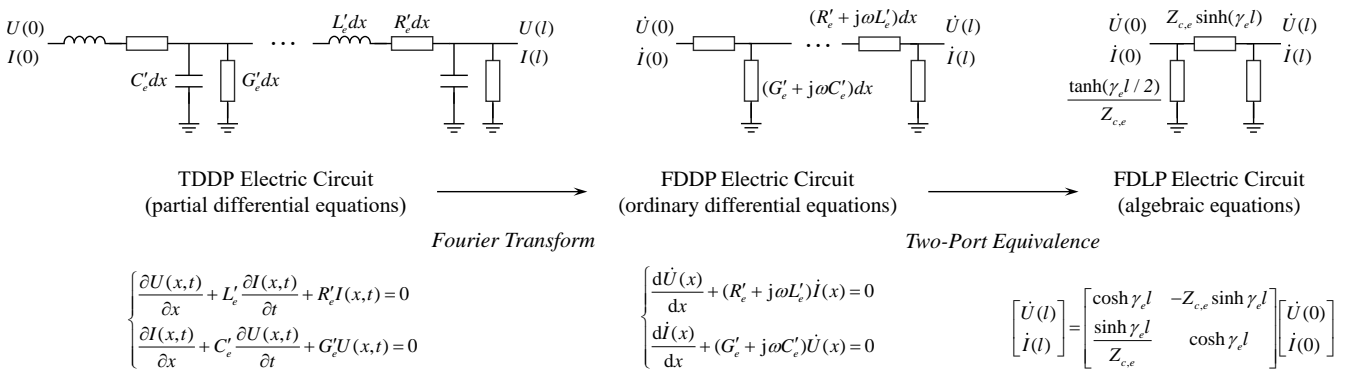

**Fig. 1** Electric circuit modeling of transmission lines in EPNs, which is essentially an algebraization procedure.

The remainder of this paper is organized as follows. In Section 2, energy circuit models, including a hydraulic circuit model for NGNs and a thermal circuit model for DHNs, are introduced. Section 3 first formulates the ECM-based OEF model and then presents the corresponding variable space projection and constraint generation algorithm. In Section 4, numerical tests are performed on two IESs with different scales to validate the proposed method and its scalability. Section 5 concludes the paper and discusses future work.

## 2. Energy Circuit Method

The distributions of voltage and current along a transmission line were initially described by the telegrapher's equations [35], which are intractable PDEs corresponding to a time-domain-distributed-parameter (TDDP) circuit. Using the Fourier transform to eliminate the temporal derivative and two-port equivalence to eliminate the spatial derivative, a frequency-domain-lumped-parameter (FDLP) circuit, i.e., the well-known pi-equivalent circuit, is ultimately derived, which yields an algebraic relationship between the voltage and current on two ends. Such a two-step transformation[†], as visualized in Fig. 1, has achieved great success in power system analysis.

Comparing pressure and temperature to voltage, and mass flow and heat flow to current, similar algebraization operations are implemented for the PDE models of pipelines in NGNs and DHNs. Because they are inspired by electric circuit modeling, these two models are named the hydraulic circuit and thermal circuit, respectively.

### 2.1 Hydraulic circuit modeling for NGNs

(1) Pipeline model

The gas flow along a pipeline obeys both the mass conservation equation and momentum conservation equation [22], as shown in (1).

$$\begin{cases} \dfrac{\partial p(x,t)}{\partial t} + \dfrac{c_s^2}{S} \dfrac{\partial m(x,t)}{\partial x} = 0 \\ \dfrac{1}{S} \dfrac{\partial m(x,t)}{\partial t} + \dfrac{\partial p(x,t)}{\partial x} + \dfrac{\lambda \rho(x,t) v^2(x,t)}{2D} + \rho(x,t) g \sin \alpha = 0 \end{cases} \quad (1)$$

Taylor-expansion linearization is adopted for the quadratic term in (1) at $v = \bar{v}$, as expressed in (2). Note that similar linearization techniques are widely used in FDM-based models, such as in [19] and [24]-[25].

$$v^2(x,t) \approx 2\bar{v}v(x,t) - \bar{v}^2 \quad (2)$$

In addition, state variables $m$, $\rho$, and $v$ satisfy the mass flow equation (3), and $p$ and $\rho$ satisfy the gas state equation (4).

$$m(x,t) = \rho(x,t) v(x,t) S \quad (3)$$

$$p(x,t) = c_s^2 \rho(x,t) \quad (4)$$

Substituting (2)-(4) into (1) yields (5), which is similar in form to the telegrapher's equations so that the TDDP hydraulic circuit of a gas pipeline is drawn as Fig. 2(a). Compared with the TDDP electric circuit, there is one less shunt conductance, which means there is no gas leakage along the transmission pipeline, and one more pressure-controlled pressure source, which means the pipeline is asymmetric at an inclined angle.

$$\begin{cases} \dfrac{\partial p(x,t)}{\partial x} + L'_g \dfrac{\partial m(x,t)}{\partial t} + R'_g m(x,t) + K'_g p(x,t) = 0 \\ \dfrac{\partial m(x,t)}{\partial x} + C'_g \dfrac{\partial p(x,t)}{\partial t} = 0 \end{cases} \quad (5)$$

where the distributed element parameters are given below:

$$\begin{cases} R'_g = \lambda \bar{v} / (AD) \\ L'_g = 1/S \\ C'_g = S / c_s^2 \\ K'_g = (2Dg \sin \alpha - \lambda \bar{v}^2) / (2Dc_s^2) \end{cases} \quad (6)$$

For the convenience of illustration, we tentatively assume that all pressure and gas flow in NGNs are sinusoidal at the same angular frequency of $\omega$ (nonsinusoidal cases will be discussed in Section 2.3). Thereby, (7) is obtained from (5) using the Fourier transform, corresponding to the FDDP hydraulic circuit shown in Fig. 2(b).

$$\begin{cases} \dfrac{d\dot{p}(x)}{dx} + (R'_g + j\omega L'_g) \dot{m}(x) + k'_g \dot{p}(x) = 0 \\ \dfrac{d\dot{m}(x)}{dx} + j\omega C'_g \dot{p}(x) = 0 \end{cases} \quad (7)$$

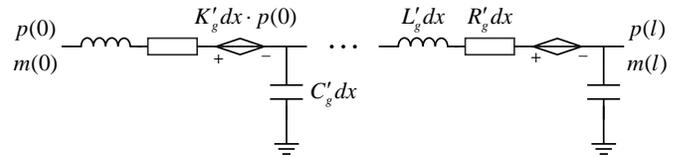

(a) TDDP Hydraulic Circuit

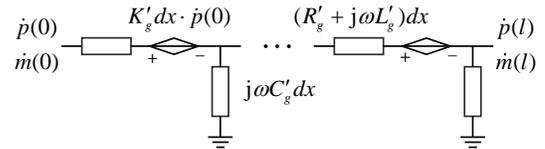

(b) FDDP Hydraulic Circuit

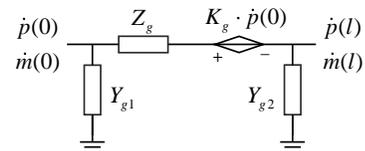

(c) FDLP Hydraulic Circuit

**Fig. 2** Hydraulic circuit modeling of pipelines in NGNs.

---
[†] We refer readers to [36] for more details.

Solving this ordinary differential equation yields a two-port network as expressed in (8), whose transmission parameters are given in (9).

$$\begin{bmatrix} \dot{p}(l) \\ \dot{m}(l) \end{bmatrix} = \begin{bmatrix} \mathcal{A}_g & \mathcal{B}_g \\ \mathcal{C}_g & \mathcal{D}_g \end{bmatrix} \begin{bmatrix} \dot{p}(0) \\ \dot{m}(0) \end{bmatrix} \quad (8)$$

$$\begin{cases} \mathcal{A}_g = \cosh\left(\sqrt{K_g'^2 + 4\gamma_g^2}\, l/2\right) e^{-K_g' l/2} - \\ \qquad \sinh\left(\sqrt{K_g'^2 + 4\gamma_g^2}\, l/2\right) K_g' e^{-K_g' l/2} / \sqrt{K_g'^2 + 4\gamma_g^2} \\ \mathcal{B}_g = -2\sinh\left(\sqrt{K_g'^2 + 4\gamma_g^2}\, l/2\right) e^{-K_g' l/2} / \sqrt{K_g'^2 + 4/Z_{c,g}^2} \\ \mathcal{C}_g = -2\sinh\left(\sqrt{K_g'^2 + 4\gamma_g^2}\, l/2\right) e^{-K_g' l/2} / \sqrt{K_g'^2 + 4 Z_{c,g}^2} \\ \mathcal{D}_g = \cosh\left(\sqrt{K_g'^2 + 4\gamma_g^2}\, l/2\right) e^{-K_g' l/2} + \\ \qquad \sinh\left(\sqrt{K_g'^2 + 4\gamma_g^2}\, l/2\right) K_g' e^{-K_g' l/2} / \sqrt{K_g'^2 + 4\gamma_g^2} \end{cases} \quad (9)$$

Generalized from that of a transmission line, $\gamma_g$ and $Z_{c,g}$ in (9) are named the propagation constant and characteristic impedance of the gas pipeline, whose definitions are

$$\begin{cases} \gamma_g = \sqrt{(R_g' + j\omega L_g') \cdot j\omega C_g'} \\ Z_{c,g} = \sqrt{(R_g' + j\omega L_g') / j\omega C_g'} \end{cases} \quad (10)$$

The two-port network (8) can be equivalently represented by the FDLP hydraulic circuit shown in Fig. 2(c), and (11) expresses the parameters of its lumped elements.

$$\begin{cases} Z_g = -\mathcal{B}_g \\ K_g = 1 - \mathcal{A}_g \mathcal{D}_g + \mathcal{B}_g \mathcal{C}_g \\ Y_{g1} = (\mathcal{A}_g \mathcal{D}_g - \mathcal{B}_g \mathcal{C}_g - \mathcal{A}_g)/\mathcal{B}_g \\ Y_{g2} = (1 - \mathcal{D}_g)/\mathcal{B}_g \end{cases} \quad (11)$$

(2) Network model

Using the FDLP hydraulic circuit, an NGN with $N_p$ pipelines is modeled as $3N_p$ branches, and these branches satisfy the following branch equation:

$$\dot{m}_b = Y_{g,b}(\dot{p}_b - K_{g,b}\dot{p}_{bf}) \quad (12)$$

where, $\dot{m}_b$ is a vector whose $i$-th entry is the gas flow on branch $i$, $\dot{p}_b$ is a vector whose $i$-th entry is the pressure difference on branch $i$, $\dot{p}_{bf}$ is a vector whose $i$-th entry is the pressure at the "from" end of branch $i$, $Y_{g,b}$ is a diagonal matrix whose $i$-th diagonal entry is the lumped admittance of branch $i$, and $K_{g,b}$ is a diagonal matrix whose $i$-th diagonal entry is the lumped controlled source parameter of branch $i$.

Then, two incidence matrices are introduced to correlate branch variables with node variables: the node-branch incidence matrix $A$, whose entry at the $i$-th row and $j$-th column is defined by (13), and the node-outflow-branch incidence matrix $A_+$, whose entry at the $i$-th row and $j$-th column is defined by (14).

$$A[i,j] = \begin{cases} 1, & \text{if branch } j \text{ flows from node } i \\ -1, & \text{if branch } j \text{ flows into node } i \\ 0, & \text{otherwise} \end{cases} \quad (13)$$

$$A_+[i,j] = \begin{cases} 1, & \text{if branch } j \text{ flows from node } i \\ 0, & \text{otherwise} \end{cases} \quad (14)$$

Using these two incidence matrices, the following topological relationships are obtained:

$$\begin{cases} A\dot{m}_b = \dot{m}_n \\ A^T \dot{p}_n = \dot{p}_b \\ A_+^T \dot{p}_n = \dot{p}_{bf} \end{cases} \quad (15)$$

where, $\dot{m}_n$ is a vector whose $i$-th entry is the gas flow injection of node $i$, and $\dot{p}_n$ is a vector whose $i$-th entry is the pressure of node $i$.

Combining the branch equation in (12) and the topological relationships in (15) yields the hydraulic circuit model of an NGN [33], as expressed in (16).

$$\dot{m}_n = (AY_{g,b}A^T - AY_{g,b}K_{g,b}A_+^T)\dot{p}_n \triangleq Y_{g,n}\dot{p}_n \quad (16)$$

where $Y_{g,n}$ is named the generalized node admittance matrix of an NGN.

### 2.2 Thermal circuit modeling for DHNs

(1) Pipeline model

The water flow in a pipeline obeys both the energy conservation equation and the enthalpy equation [23], as expressed in (17).

$$\begin{cases} c_p \rho S \dfrac{\partial T(x,t)}{\partial t} + c_p m \dfrac{\partial T(x,t)}{\partial x} + \mu T(x,t) = 0 \\ h(x,t) = c_p m T(x,t) \end{cases} \quad (17)$$

In this paper, we focus on the quality-regulation mode of DHN operation that is commonly adopted in Russia, Nordic countries [37], and northern China, in which the water flow $m$ is a preset constant. Moreover, the density $\rho$ is also a constant considering that water is incompressible. Note that $T$ in (17) is the relative temperature after subtracting the ambient temperature.

(17) is arranged into the same form as the telegrapher's equation, as expressed in (18), which gives the TDDP thermal circuit of a heat pipeline in Fig. 3(a).

$$\begin{cases} \dfrac{\partial T(x,t)}{\partial x} + L_h' \dfrac{\partial h(x,t)}{\partial t} + R_h' h(x,t) = 0 \\ \dfrac{\partial h(x,t)}{\partial x} + C_h' \dfrac{\partial T(x,t)}{\partial t} + G_h' T(x,t) = 0 \end{cases} \quad (18)$$

where the distributed element parameters are given below:

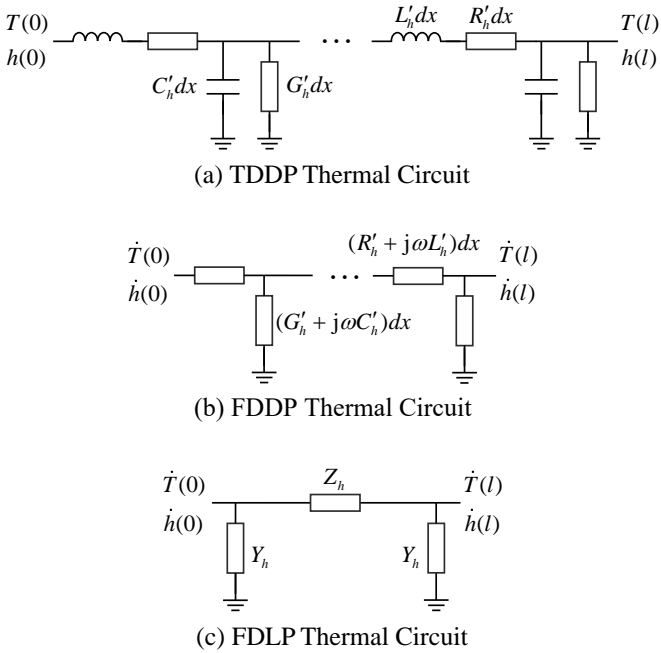

(a) TDDP Thermal Circuit

(b) FDDP Thermal Circuit

(c) FDLP Thermal Circuit

**Fig. 3** Thermal circuit modeling of pipelines in DHNs.

$$\begin{cases} R'_h = \mu / c_p^2 / m^2 \\ L'_h = \rho S / c_p / m^2 \\ G'_h = \mu \\ C'_h = c_p \rho S \end{cases} \quad (19)$$

Similarly, we tentatively assume that all temperature and heat flows in DHNs are sinusoidal at the same angular frequency of $\omega$. Therefore, applying the Fourier transform to (18) yields (20), which gives the FDDP thermal circuit of a heat pipeline, as shown in Fig. 3(b).

$$\begin{cases} \dfrac{d\dot{T}(x)}{dx} + (R'_h + j\omega L'_h)\dot{h}(x) = 0 \\ \dfrac{d\dot{h}(x)}{dx} + (G'_h + j\omega C'_h)\dot{T}(x) = 0 \end{cases} \quad (20)$$

The solution to (20) yields a two-port network as expressed in (21), whose transmission parameters are given in (22). Because a heat pipeline is symmetric, its two-port network is reciprocal so that the transmission parameters satisfy $\mathcal{A}_h \mathcal{D}_h - \mathcal{B}_h \mathcal{C}_h = 1$ and $\mathcal{A}_h = \mathcal{D}_h$.

$$\begin{bmatrix} \dot{T}(l) \\ \dot{h}(l) \end{bmatrix} = \begin{bmatrix} \mathcal{A}_h & \mathcal{B}_h \\ \mathcal{C}_h & \mathcal{D}_h \end{bmatrix} \begin{bmatrix} \dot{T}(0) \\ \dot{h}(0) \end{bmatrix} \quad (21)$$

$$\begin{cases} \mathcal{A}_h = \cosh(\gamma_h l) \\ \mathcal{B}_h = -\sinh(\gamma_h l) \cdot Z_{c,h} \\ \mathcal{C}_h = -\sinh(\gamma_h l) / Z_{c,h} \\ \mathcal{D}_h = \cosh(\gamma_h l) \end{cases} \quad (22)$$

where $Z_{c,h}$ and $\gamma_h$ are named the characteristic impedance and propagation constant of a heat pipeline, respectively:

$$\begin{cases} Z_{c,h} = \sqrt{(R'_h + j\omega L'_h)/(G'_h + j\omega C'_h)} \\ \gamma_h = \sqrt{(R'_h + j\omega L'_h) \cdot (G'_h + j\omega C'_h)} \end{cases} \quad (23)$$

The pi-equivalent FDLP thermal circuit for this two-port network is shown in Fig. 3(c). Using the reciprocal condition, its lumped-element parameters have a simpler form, as expressed in (24).

$$\begin{cases} Z_h = -\mathcal{B}_h \\ Y_h = (1 - \mathcal{A}_h)/\mathcal{B}_h \end{cases} \quad (24)$$

Substituting the enthalpy equation $\dot{h} = c_p m \dot{T}$ into (21), the two-port network is further reduced to a one-port network whose input and output are both heat flows:

$$\dot{h}(l) = e^{-\gamma l} \cdot \dot{h}(0) \triangleq \phi \cdot \dot{h}(0) \quad (25)$$

where $\phi$ is named the heat transmission factor.

(2) Network model

Based on (25), a DHN with $N_p$ pipelines can be modeled as $N_p$ branches, and these branches satisfy the branch equation in (26).

$$\dot{\boldsymbol{h}}_{bt} = \boldsymbol{\phi}_b \dot{\boldsymbol{h}}_{bf} \quad (26)$$

where, $\dot{\boldsymbol{h}}_{bf} / \dot{\boldsymbol{h}}_{bt}$ is a vector whose $i$-th entry is the heat flow at the "from"/"to" end of branch $i$, and $\boldsymbol{\phi}_b$ is a diagonal matrix whose $i$-th diagonal entry is the heat transmission factor of branch $i$.

In addition to $\boldsymbol{A}$ and $\boldsymbol{A}_+$, another node-inflow-branch incidence matrix $\boldsymbol{A}_-$ is introduced, whose entry at the $i$-th row and $j$-th column is defined in (27).

$$\boldsymbol{A}_-[i,j] = \begin{cases} 1, & \text{if branch } j \text{ flows into node } i \\ 0, & \text{otherwise} \end{cases} \quad (27)$$

Using these incidence matrices, the confluence and divergence relationships of heat flows are given in (28).

$$\begin{cases} \dot{\boldsymbol{h}}_\Sigma = \boldsymbol{A}_- \dot{\boldsymbol{h}}_{bt} + \dot{\boldsymbol{h}}_n \\ \dot{\boldsymbol{h}}_{bf} = \tilde{\boldsymbol{A}}_+^T \dot{\boldsymbol{h}}_\Sigma \end{cases} \quad (28)$$

where, $\dot{\boldsymbol{h}}_\Sigma$ is a vector whose $i$-th entry is the total heat flow passing node $i$, $\dot{\boldsymbol{h}}_n$ is a vector whose $i$-th entry is the heat flow injection of node $i$, and $\tilde{\boldsymbol{A}}_+$ is defined in (29) whose nonzero entry at the $i$-th row and $j$-th column is the proportion of the water flow on branch $j$ to the total water flow passing node $i$.

$$\tilde{\boldsymbol{A}}_+ = \text{diag}^{-1}(\boldsymbol{A}_+ \boldsymbol{m}_{b,h}) \boldsymbol{A}_+ \text{diag}(\boldsymbol{m}_{b,h}) \quad (29)$$

where $\boldsymbol{m}_{b,h}$ is a vector whose $i$-th entry is the water flow on branch *i*.

Combining the branch equation in (26) and the topological relationships in (28) yields the thermal circuit model of a DHN [34], as expressed in (30).

$$\begin{aligned}\tilde{\boldsymbol{A}}_+^T \dot{\boldsymbol{h}}_n &= (\boldsymbol{I} - \tilde{\boldsymbol{A}}_+^T \boldsymbol{A}_- \boldsymbol{\eta}_b) \dot{\boldsymbol{h}}_{bf} \\ &= (\boldsymbol{I} - \tilde{\boldsymbol{A}}_+^T \boldsymbol{A}_- \boldsymbol{\eta}_b) \text{diag}(c_p \boldsymbol{m}_{b,h}) \dot{\boldsymbol{T}}_{bf} \\ &\triangleq \boldsymbol{Y}_{h,b} \dot{\boldsymbol{T}}_{bf}\end{aligned} \quad (30)$$

where, $\boldsymbol{I}$ is an identity matrix of order $M$, $\dot{\boldsymbol{T}}_{bf}$ is a vector whose *i*-th entry is the temperature at the "from" end of branch *i*, and $\boldsymbol{Y}_{h,b}$ is named the generalized branch admittance matrix of a DHN.

From (30), the node temperature in a DHN is obtained:

$$\dot{\boldsymbol{T}}_n = \tilde{\boldsymbol{A}}_+ \dot{\boldsymbol{T}}_{bf} = \tilde{\boldsymbol{A}}_+ \boldsymbol{Y}_{h,b}^{-1} \tilde{\boldsymbol{A}}_+^T \dot{\boldsymbol{h}}_n \triangleq \boldsymbol{Z}_{h,n} \dot{\boldsymbol{h}}_n \quad (31)$$

where, $\dot{\boldsymbol{T}}_n$ is a vector whose *i*-th entry is the temperature of node *i* and $\boldsymbol{Z}_{h,n}$ is named the generalized node impedance matrix of a DHN.

*2.3 More discussions*

(1) Regarding nonsinusoidal cases

Both aforementioned energy circuit models for NGNs and DHNs are obtained under the assumption of sinusoidal variables at the same frequency. For nonsinusoidal cases, there are multiple sinusoidal components at different frequencies after the Fourier transform. According to the superposition theorem, the derived energy circuit equations still hold for each independent frequency component. Therefore, the energy circuit model for a general NGN/DHN contains a series of circuits to describe nonsinusoidal dynamic processes. Note that these circuits have the same time-domain parameters (i.e., *R*, *L*, *G*, *C*) but not the same frequency-domain parameters (i.e., *Z*, *Y*), since they have different angular frequencies. For better clarification, an instance with three frequency components is illustrated in Fig. 4, in which *I* represents gas flow or heat flow and *U* represents pressure or temperature.

(2) Regarding determining a particular solution

To determine a particular solution from the PDE model of an NGN/DHN, both a boundary condition and an initial condition should be specified, because the current state is uniquely determined by the previous state and current boundary, as shown in Fig. 5(a). For example, by giving the pressure distribution in an NGN at the initial moment (*initial condition*) and the supplies and loads of natural gas along a period of time (*boundary condition*), pressure distributions in this time interval (*particular solution*) are obtained by solving the corresponding PDE model.

In the ECM, the boundary condition is decomposed into several sinusoidal components by the discrete Fourier transform, which implicitly performs a periodic extension, as shown in Fig. 5(b). Thus, the solution given by the ECM is a general solution without specifying the initial condition. If the state at the initial moment of this general solution is largely

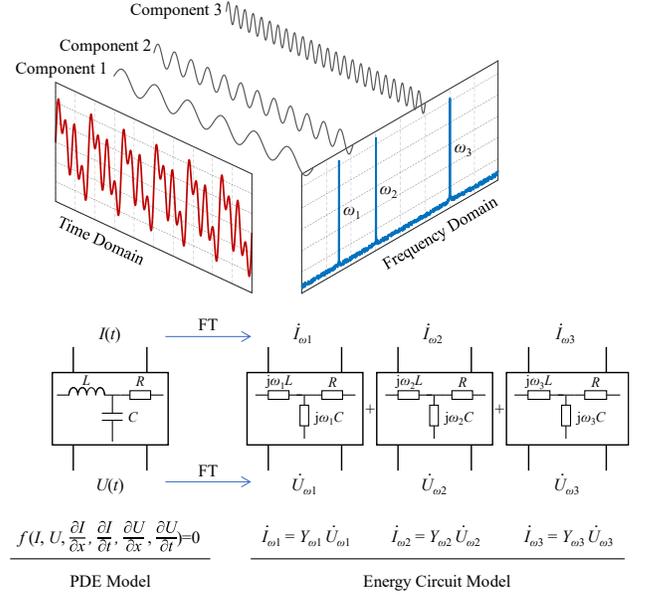

**Fig. 4** Modeling of nonsinusoidal cases using multiple energy circuits at different frequencies.

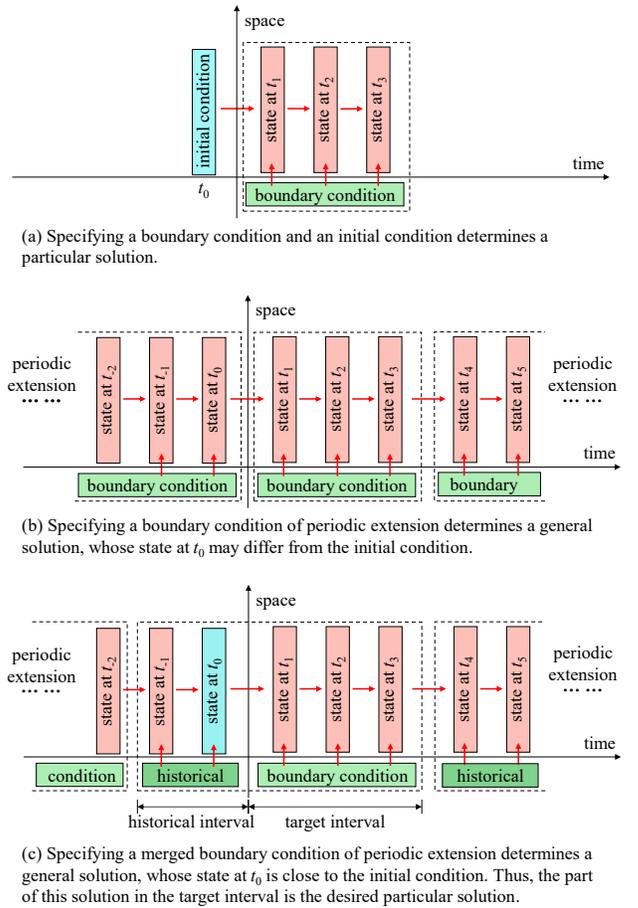

(a) Specifying a boundary condition and an initial condition determines a particular solution.

(b) Specifying a boundary condition of periodic extension determines a general solution, whose state at $t_0$ may differ from the initial condition.

(c) Specifying a merged boundary condition of periodic extension determines a general solution, whose state at $t_0$ is close to the initial condition. Thus, the part of this solution in the target interval is the desired particular solution.

**Fig. 5** Determination of a PDE solution using different conditions.

different from the initial condition, then this solution is not the desired particular solution.

For a continuously operating physical system, its state at a moment is not random, but is a result of all boundary conditions before this moment. Considering the response decay, a boundary condition of finite length before this moment is able to produce a similar state to the state of this moment. Therefore, splicing a historical boundary condition before the given boundary condition and performing the ECM on this merged boundary condition return a general solution whose state at the initial moment is close to the initial condition, as illustrated in Fig. 5(c). Consequently, the part of this general solution in the target interval is exactly the desired particular solution. The essence of such an approximation is the utilization of a historical boundary condition as a surrogate initial condition.

## 3. ECM-based Optimal Energy Flow Model

In this section, the dynamic OEF model based on the ECM is formulated for an IES that consists of an EPN, an NGN, and a DHN, in which devices of coal-fired thermal power units (TPUs), natural gas units (NGUs), combined heat and power (CHP) units, wind turbines, gas-fired boilers, heat pumps, and gas wells are considered. An $N_{ht}$-point historical boundary condition, whose indices are $\Omega_{ht} = \{0, 1, \ldots, N_{ht} - 1\}$, is used to surrogate the initial condition, and $N_{dt}$ dispatch periods, whose indices are $\Omega_{dt} = \{N_{ht}, N_{ht} + 1, \ldots, N_{ht} + N_{dt} - 1\}$, are considered for the optimization of device schedules. According to the theory of Fourier series, such an $(N_{ht} + N_{dt})$-point time-domain series has $N_f$ frequency components, where $N_f = 1 + \lfloor (N_{ht} + N_{dt})/2 \rfloor$. Its physical meaning is that the superposition of $N_f$ sinusoidal series can exactly fit an arbitrary $(N_{ht} + N_{dt})$-point nonsinusoidal series. We denote the indices of frequency components as $\Omega_f = \{0, 1, \ldots, N_f - 1\}$.

Afterward, the variable space projection and constraint generation algorithm are supplemented to obtain a more compact ECM-based OEF model.

### 3.1 Baseline model

(1) Framework

The framework of the ECM-based OEF model is illustrated in Fig. 6. The decision variables are categorized from two dimensions: time-domain variables or frequency-domain variables, and controllable variables (i.e., device outputs) or monitored variables (i.e., pressure and temperature), which divide the variable space into four quadrants. The conventional constraints are constructed using time-domain variables, including output, ramping, coupling, and balance constraints for controllable variables and security constraints for monitored variables. To incorporate the energy circuit models of the NGN and the DHN, time-frequency conversion constraints are introduced to couple the time-domain variables and frequency-domain variables, so that the energy circuit constraints of the controllable variables and monitored variables in the frequency domain indirectly influence these variables in the time domain. Moreover, some other constraints regarding the freedom degree and historical boundary are constructed for frequency-domain controllable variables. The main objective of OEF is to minimize the operating cost that is formulated by the time-domain controllable variables. With the same optimality, another auxiliary objective to smooth dispatch curves is achieved by imposing a sufficiently small penalty on the high-frequency component magnitudes of the frequency-domain controllable variables.

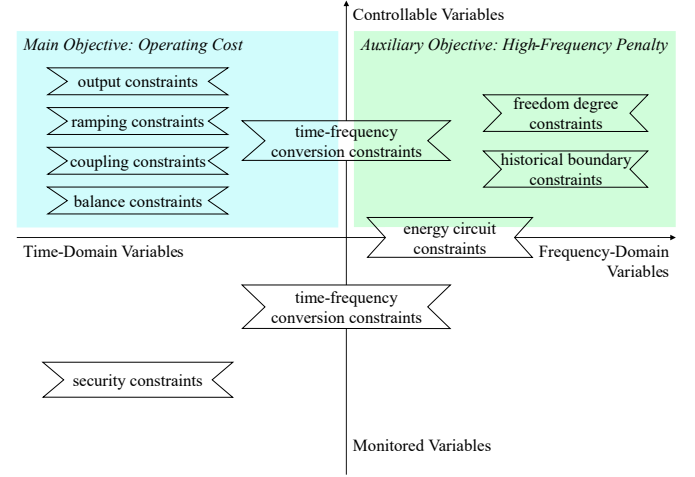

**Fig. 6** Framework of the ECM-based OEF model.

These variables, constraints, and objectives are formulated mathematically as follows.

(2) Decision variables and constants

The decision variables involved in the ECM-based OEF model are listed in Table 1. Note that each frequency-domain variable is a complex variable that contains two real variables of the real part and imaginary part.

**Table 1** Decision variables in the ECM-based OEF model

| | |
|---|---|
| i) time-domain controllable variables ($\tau \in \Omega_{dt}$) | |
| $P_{tpu,i}^{(\tau)}$ | power production of $i$-th TPU at time $\tau$ |
| $P_{ngu,i}^{(\tau)}$ | power production of $i$-th NGU at time $\tau$ |
| $P_{chp,i}^{(\tau)}$ | power production of $i$-th CHP unit at time $\tau$ |
| $P_{wt,i}^{(\tau)}$ | power production of $i$-th wind turbine at time $\tau$ |
| $P_{hp,i}^{(\tau)}$ | power consumption of $i$-th heat pump at time $\tau$ |
| $m_{gw,i}^{(\tau)}$ | gas production of $i$-th gas well at time $\tau$ |
| $m_{ngu,i}^{(\tau)}$ | gas consumption of $i$-th NGU at time $\tau$ |
| $m_{gb,i}^{(\tau)}$ | gas consumption of $i$-th gas-fired boiler at time $\tau$ |
| $h_{chp,i}^{(\tau)}$ | heat production of $i$-th CHP unit at time $\tau$ |
| $h_{gb,i}^{(\tau)}$ | heat production of $i$-th gas-fired boiler at time $\tau$ |
| $h_{hp,i}^{(\tau)}$ | heat production of $i$-th heat pump at time $\tau$ |
| ii) time-domain monitored variables ($\tau \in \Omega_{dt}$) | |
| $p_{n,i}^{(\tau)}$ | pressure of $i$-th NGN node at time $\tau$ |
| $T_{n,i}^{(\tau)}$ | temperature of $i$-th DHN node at time $\tau$ |
| iii) frequency-domain controllable variables ($\kappa \in \Omega_f$) | |
| $\dot{m}_{gw,i}^{(\kappa)}$ | $\kappa$-th frequency component of $\{m_{gw,i}^{(\tau)} \mid \tau \in \Omega_{ht} \cup \Omega_{dt}\}$ |
| $\dot{m}_{ngu,i}^{(\kappa)}$ | $\kappa$-th frequency component of $\{m_{ngu,i}^{(\tau)} \mid \tau \in \Omega_{ht} \cup \Omega_{dt}\}$ |
| $\dot{m}_{gb,i}^{(\kappa)}$ | $\kappa$-th frequency component of $\{m_{gb,i}^{(\tau)} \mid \tau \in \Omega_{ht} \cup \Omega_{dt}\}$ |
| $\dot{h}_{chp,i}^{(\kappa)}$ | $\kappa$-th frequency component of $\{h_{chp,i}^{(\tau)} \mid \tau \in \Omega_{ht} \cup \Omega_{dt}\}$ |
| $\dot{h}_{gb,i}^{(\kappa)}$ | $\kappa$-th frequency component of $\{h_{gb,i}^{(\tau)} \mid \tau \in \Omega_{ht} \cup \Omega_{dt}\}$ |
| $\dot{h}_{hp,i}^{(\kappa)}$ | $\kappa$-th frequency component of $\{h_{hp,i}^{(\tau)} \mid \tau \in \Omega_{ht} \cup \Omega_{dt}\}$ |

iv) frequency-domain monitored variables ($\kappa \in \Omega_f$)

| | |
|---|---|
| $\dot{p}_{n,i}^{(\kappa)}$ | $\kappa$-th frequency component of $\{p_{n,i}^{(\tau)} \mid \tau \in \Omega_{ht} \cup \Omega_{dt}\}$ |
| $\dot{T}_{n,i}^{(\kappa)}$ | $\kappa$-th frequency component of $\{T_{n,i}^{(\tau)} \mid \tau \in \Omega_{ht} \cup \Omega_{dt}\}$ |

The constants for the optimization boundaries involved in the ECM-based OEF model are listed in Table 2.

**Table 2** Constants in the ECM-based OEF model

| | |
|---|---|
| i) renewable energy and loads | |
| $P_{wt,i}^{ub,(\tau)}$ | available wind power of $i$-th wind turbine at time $\tau$ ($\tau \in \Omega_{dt}$) |
| $P_{ld,i}^{(\tau)}$ | power load of $i$-th EPN bus at time $\tau$ ($\tau \in \Omega_{dt}$) |
| $m_{ld,i}^{(\tau)}$ | gas load of $i$-th NGN node at time $\tau$ ($\tau \in \Omega_{ht} \cup \Omega_{dt}$) |
| $h_{ld,i}^{(\tau)}$ | heat load of $i$-th DHN node at time $\tau$ ($\tau \in \Omega_{ht} \cup \Omega_{dt}$) |
| $\dot{m}_{ld,i}^{(\kappa)}$ | $\kappa$-th frequency component of $\{m_{ld,i}^{(\tau)} \mid \tau \in \Omega_{ht} \cup \Omega_{dt}\}$ |
| $\dot{h}_{ld,i}^{(\kappa)}$ | $\kappa$-th frequency component of $\{h_{ld,i}^{(\tau)} \mid \tau \in \Omega_{ht} \cup \Omega_{dt}\}$ |
| ii) security bounds | |
| $P_{l,i}^{ub}$ | upper bound of $i$-th transmission line power |
| $p_{n,i}^{ub}$, $p_{n,i}^{lb}$ | upper and lower bounds of $i$-th NGN node pressure |
| $T_{n,i}^{ub}$, $T_{n,i}^{lb}$ | upper and lower bounds of $i$-th DHN node temperature |
| iii) network parameters | |
| **PTDF** | power transform distribution factor matrix of the EPN |
| $\boldsymbol{Y}_{g,n}^{(\kappa)}$ | generalized node admittance matrix of the NGN at $\kappa$-th frequency component |
| $\tilde{\boldsymbol{A}}_{+}$ | node-outflow-branch incidence matrix of the DHN |
| $\boldsymbol{Y}_{h,b}^{(\kappa)}$ | generalized branch admittance matrix of the DHN at $\kappa$-th frequency component |
| iv) device parameters | |
| $r_{ngu,i}$ | gas-to-power ratio of $i$-th NGU |
| $r_{chp,i}$ | heat-to-power ratio of $i$-th CHP unit |
| $r_{gb,i}$ | gas-to-heat ratio of $i$-th gas-fired boiler |
| $r_{hp,i}$ | power-to-heat ratio of $i$-th heat pump |
| $P_{tpu,i}^{ub}$, $P_{ngu,i}^{ub}$, $P_{chp,i}^{ub}$, $m_{gw,i}^{ub}$, $h_{gb,i}^{ub}$, and $h_{hp,i}^{ub}$ are upper bounds of device outputs. | |
| $P_{tpu,i}^{lb}$, $P_{ngu,i}^{lb}$, $P_{chp,i}^{lb}$, $m_{gw,i}^{lb}$, $h_{gb,i}^{lb}$, and $h_{hp,i}^{lb}$ are lower bounds of device outputs. | |
| $P_{tpu,i}^{rampU}$, $P_{ngu,i}^{rampU}$, $P_{chp,i}^{rampU}$, $m_{gw,i}^{rampU}$, $h_{gb,i}^{rampU}$, and $h_{hp,i}^{rampU}$ are up ramping limits of device outputs. | |
| $P_{tpu,i}^{rampD}$, $P_{ngu,i}^{rampD}$, $P_{chp,i}^{rampD}$, $m_{gw,i}^{rampD}$, $h_{gb,i}^{rampD}$, and $h_{hp,i}^{rampD}$ are down ramping limits of device outputs. | |
| $u_{tpu,2,i}$, $u_{tpu,1,i}$, $u_{tpu,0,i}$, $u_{chp,p2,i}$, $u_{chp,p1,i}$, $u_{chp,p0,i}$, $u_{chp,h2,i}$, $u_{chp,h1,i}$, $u_{chp,h0,i}$, $u_{gw,2,i}$, $u_{gw,1,i}$, and $u_{gw,0,i}$ are cost coefficients of devices. | |

The above constants are mostly input by users, except $\dot{m}_{ld,i}^{(\kappa)}$ and $\dot{h}_{ld,i}^{(\kappa)}$ are calculated by (32) using $m_{ld,i}^{(\tau)}$ and $h_{ld,i}^{(\tau)}$.

$$\begin{cases} \dot{m}_{ld,i}^{(\kappa)} = \dfrac{\upsilon(\kappa)}{N_{ht}+N_{dt}} \sum_{\tau=0}^{N_{ht}+N_{dt}-1} m_{ld,i}^{(\tau)} e^{-\frac{j2\pi\kappa\tau}{N_{ht}+N_{dt}}} \\ \dot{h}_{ld,i}^{(\kappa)} = \dfrac{\upsilon(\kappa)}{N_{ht}+N_{dt}} \sum_{\tau=0}^{N_{ht}+N_{dt}-1} h_{ld,i}^{(\tau)} e^{-\frac{j2\pi\kappa\tau}{N_{ht}+N_{dt}}} \end{cases} \quad (32)$$

where

$$\upsilon(\kappa) = \begin{cases} 1, & \text{if } \kappa = 0 \text{ or } \kappa \times 2 = N_{td} \\ 2, & \text{otherwise} \end{cases} \quad (33)$$

Combining these variables and constants, we have the vectors of the time-domain power injection of the EPN nodes, the frequency-domain gas injection of the NGN nodes, and the frequency-domain heat injection of the DHN nodes as expressed in (34)-(36), respectively.

$$\boldsymbol{P}_n^{(\tau)}[i] = \sum_{j\in i} P_{tpu,j}^{(\tau)} + \sum_{j\in i} P_{ngu,j}^{(\tau)} + \sum_{j\in i} P_{chp,j}^{(\tau)} + \sum_{j\in i} P_{wt,j}^{(\tau)} \\ - \sum_{j\in i} P_{hp,j}^{(\tau)} - P_{ld,i}^{(\tau)} \quad (34)$$

$$\dot{\boldsymbol{m}}_n^{(\kappa)}[i] = \sum_{j\in i} \dot{m}_{gw,j}^{(\kappa)} - \sum_{j\in i} \dot{m}_{ngu,j}^{(\kappa)} - \sum_{j\in i} \dot{m}_{gb,j}^{(\kappa)} - \dot{m}_{ld,i}^{(\kappa)} \quad (35)$$

$$\dot{\boldsymbol{h}}_n^{(\kappa)}[i] = \sum_{j\in i} \dot{h}_{chp,j}^{(\kappa)} + \sum_{j\in i} \dot{h}_{gb,j}^{(\kappa)} + \sum_{j\in i} \dot{h}_{hp,j}^{(\kappa)} - \dot{h}_{ld,i}^{(\kappa)} \quad (36)$$

where $j \in i$ denotes device $j$ that connects to bus/node $i$.

(3) Constraints

i) output constraints of devices

$$\begin{cases} 0 \le P_{wt,i}^{(\tau)} \le P_{wt,i}^{ub,(\tau)} & \forall i \in \Omega_{wt}, \tau \in \Omega_{dt} \\ P_{tpu,i}^{lb} \le P_{tpu,i}^{(\tau)} \le P_{tpu,i}^{ub} & \forall i \in \Omega_{tpu}, \tau \in \Omega_{dt} \\ P_{ngu,i}^{lb} \le P_{ngu,i}^{(\tau)} \le P_{ngu,i}^{ub} & \forall i \in \Omega_{ngu}, \tau \in \Omega_{dt} \\ P_{chp,i}^{lb} \le P_{chp,i}^{(\tau)} \le P_{chp,i}^{ub} & \forall i \in \Omega_{chp}, \tau \in \Omega_{dt} \\ m_{gw,i}^{lb} \le m_{gw,i}^{(\tau)} \le m_{gw,i}^{ub} & \forall i \in \Omega_{gw}, \tau \in \Omega_{dt} \\ h_{gb,i}^{lb} \le h_{gb,i}^{(\tau)} \le h_{gb,i}^{ub} & \forall i \in \Omega_{gb}, \tau \in \Omega_{dt} \\ h_{hp,i}^{lb} \le h_{hp,i}^{(\tau)} \le h_{hp,i}^{ub} & \forall i \in \Omega_{hp}, \tau \in \Omega_{dt} \end{cases} \quad (37)$$

ii) ramping constraints of devices

$$\begin{cases} -P_{tpu,i}^{rampD} \le P_{tpu,i}^{(\tau)} - P_{tpu,i}^{(\tau-1)} \le P_{tpu,i}^{rampU} & \forall i \in \Omega_{tpu}, \tau \in \Omega_{dt} \\ -P_{ngu,i}^{rampD} \le P_{ngu,i}^{(\tau)} - P_{ngu,i}^{(\tau-1)} \le P_{ngu,i}^{rampU} & \forall i \in \Omega_{ngu}, \tau \in \Omega_{dt} \\ -P_{chp,i}^{rampD} \le P_{chp,i}^{(\tau)} - P_{chp,i}^{(\tau-1)} \le P_{chp,i}^{rampU} & \forall i \in \Omega_{chp}, \tau \in \Omega_{dt} \\ -m_{gw,i}^{rampD} \le m_{gw,i}^{(\tau)} - m_{gw,i}^{(\tau-1)} \le m_{gw,i}^{rampU} & \forall i \in \Omega_{gw}, \tau \in \Omega_{dt} \\ -h_{gb,i}^{rampD} \le h_{gb,i}^{(\tau)} - h_{gb,i}^{(\tau-1)} \le h_{gb,i}^{rampU} & \forall i \in \Omega_{gb}, \tau \in \Omega_{dt} \\ -h_{hp,i}^{rampD} \le h_{hp,i}^{(\tau)} - h_{hp,i}^{(\tau-1)} \le h_{hp,i}^{rampU} & \forall i \in \Omega_{hp}, \tau \in \Omega_{dt} \end{cases} \quad (38)$$

iii) coupling constraints of devices

$$\begin{cases} r_{ngu,i} m_{ngu,i}^{(\tau)} = P_{ngu,i}^{(\tau)} & \forall i \in \Omega_{ngu}, \tau \in \Omega_{dt} \\ r_{chp,i} h_{chp,i}^{(\tau)} = P_{chp,i}^{(\tau)} & \forall i \in \Omega_{chp}, \tau \in \Omega_{dt} \\ r_{gb,i} m_{gb,i}^{(\tau)} = h_{gb,i}^{(\tau)} & \forall i \in \Omega_{gb}, \tau \in \Omega_{dt} \\ r_{hp,i} P_{hp,i}^{(\tau)} = h_{hp,i}^{(\tau)} & \forall i \in \Omega_{hp}, \tau \in \Omega_{dt} \end{cases} \quad (39)$$

iv) frequency-domain freedom degree constraints of devices

The imaginary part of the zero-frequency component should be zero. In addition, to ensure that the mapping relationship between the time domain and frequency domain is unique, the time-domain series and frequency-domain series should have the same number of real variables. Thus, if $N_{ht} + N_{dt}$ is even, the imaginary part of the last frequency component should also be zero.

$$\begin{cases} \text{Im}(\dot{m}_{gw,i}^{(\kappa)}) = 0 & \forall i \in \Omega_{gw}, \kappa \in \{0,(N_{ht}+N_{dt})/2\} \cap \mathbb{Z} \\ \text{Im}(\dot{m}_{ngu,i}^{(\kappa)}) = 0 & \forall i \in \Omega_{ngu}, \kappa \in \{0,(N_{ht}+N_{dt})/2\} \cap \mathbb{Z} \\ \text{Im}(\dot{h}_{chp,i}^{(\kappa)}) = 0 & \forall i \in \Omega_{chp}, \kappa \in \{0,(N_{ht}+N_{dt})/2\} \cap \mathbb{Z} \\ \text{Im}(\dot{h}_{gb,i}^{(\kappa)}) = 0 & \forall i \in \Omega_{gb}, \kappa \in \{0,(N_{ht}+N_{dt})/2\} \cap \mathbb{Z} \\ \text{Im}(\dot{h}_{hp,i}^{(\kappa)}) = 0 & \forall i \in \Omega_{hp}, \kappa \in \{0,(N_{ht}+N_{dt})/2\} \cap \mathbb{Z} \end{cases} \quad (40)$$

v) supply-demand balance constraints of networks

The real-time power balance between the supply and demand in the EPN should be maintained, as expressed in (41).

$$\sum_{i \in \Omega_{tpu}} P_{tpu,i}^{(\tau)} + \sum_{i \in \Omega_{ngu}} P_{ngu,i}^{(\tau)} + \sum_{i \in \Omega_{chp}} P_{chp,i}^{(\tau)} + \sum_{i \in \Omega_{wt}} P_{wt,i}^{(\tau)} = \sum_{i \in \Omega_{e,n}} P_{ld,i}^{(\tau)} + \sum_{i \in \Omega_{hp}} P_{hp,i}^{(\tau)} \quad \forall \tau \in \Omega_{dt} \quad (41)$$

Although temporary supply-demand mismatch is allowed in the NGN and DHN, a balance between the total supply and total demand along all dispatch periods should be maintained for sustainable operation, as expressed in (42). For the DHN, the heat loss during transmission cannot be predicted, so we use an inequality constraint to relax the heat loss.

$$\begin{cases} \sum_{\tau \in \Omega_{dt}} \left( \sum_{i \in \Omega_{gw}} m_{gw,i}^{(\tau)} - \sum_{i \in \Omega_{ngu}} m_{ngu,i}^{(\tau)} - \sum_{i \in \Omega_{gb}} m_{gb,i}^{(\tau)} - \sum_{i \in \Omega_{g,n}} m_{ld,i}^{(\tau)} \right) = 0 \\ \sum_{\tau \in \Omega_{dt}} \left( \sum_{i \in \Omega_{chp}} h_{chp,i}^{(\tau)} + \sum_{i \in \Omega_{gb}} h_{gb,i}^{(\tau)} + \sum_{i \in \Omega_{hp}} h_{hp,i}^{(\tau)} - \sum_{i \in \Omega_{h,n}} h_{ld,i}^{(\tau)} \right) \geq 0 \end{cases} \quad (42)$$

vi) security constraints of networks

For the EPN, the power of each transmission line should not exceed its capacity, as expressed in (43). For the NGN and DHN, the node pressure and node temperature should not exceed the secure range, as expressed in (44).

$$-P_{l,i}^{ub} \leq \boldsymbol{PTDF}[i,:] \cdot \boldsymbol{P}_n^{(\tau)} \leq P_{l,i}^{ub} \quad \forall i \in \Omega_{e,l}, \tau \in \Omega_{dt} \quad (43)$$

$$\begin{cases} p_{n,i}^{lb} \leq p_{n,i}^{(\tau)} \leq p_{n,i}^{ub} & \forall i \in \Omega_{g,n}, \tau \in \Omega_{dt} \\ T_{n,i}^{lb} \leq T_{n,i}^{(\tau)} \leq T_{n,i}^{ub} & \forall i \in \Omega_{h,n}, \tau \in \Omega_{dt} \end{cases} \quad (44)$$

vii) energy circuit constraints of networks

As derived in Section 2, the node pressure and gas injection in the NGN satisfy (45), and the node temperature and heat injection satisfy (46). Note that we use (46) rather than $\dot{T}_n^{(\kappa)} = Z_{h,n}^{(\kappa)} \dot{h}_n^{(\kappa)}$ for DHN modeling, because the admittance matrix is far sparser than the impedance matrix, which is significantly beneficial for the solving efficiency.

$$Y_{g,n}^{(\kappa)} \dot{p}_n^{(\kappa)} = \dot{m}_n^{(\kappa)} \quad \forall \kappa \in \Omega_f \quad (45)$$

$$\begin{cases} \tilde{A}_+^T \dot{h}_n^{(\kappa)} = Y_{h,b}^{(\kappa)} \dot{T}_{bf}^{(\kappa)} \\ \dot{T}_n^{(\kappa)} = \tilde{A}_+ \dot{T}_{bf}^{(\kappa)} \end{cases} \quad \forall \kappa \in \Omega_f \quad (46)$$

where $\dot{\boldsymbol{p}}_n^{(\kappa)} = [\dot{p}_{n,1}^{(\kappa)}, \dot{p}_{n,2}^{(\kappa)}, \cdots \dot{p}_{n,N_{g,n}}^{(\kappa)}]^T$ and $\dot{\boldsymbol{T}}_n^{(\kappa)} = [\dot{T}_{n,1}^{(\kappa)}, \dot{T}_{n,2}^{(\kappa)}, \cdots, \dot{T}_{n,N_{h,n}}^{(\kappa)}]^T$.

viii) time-frequency conversion constraints of both devices and networks

The inverse discrete Fourier transform (IDFT) correlates the time-domain variables and frequency-domain variables, as expressed in (47) for controllable variables and (48) for monitored variables. Note that historical boundary constraints are also included in (47), which reflects that their right-hand sides for $\tau \in \Omega_{ht}$ are constants.

$$\begin{cases} \sum_{\kappa \in \Omega_f} \text{Re}(\dot{m}_{gw,i}^{(\kappa)} e^{j2\pi\kappa\tau/(N_{ht}+N_{dt})}) = m_{gw,i}^{(\tau)} & \forall i \in \Omega_{gw} \\ \sum_{\kappa \in \Omega_f} \text{Re}(\dot{m}_{ngu,i}^{(\kappa)} e^{j2\pi\kappa\tau/(N_{ht}+N_{dt})}) = m_{ngu,i}^{(\tau)} & \forall i \in \Omega_{ngu} \\ \sum_{\kappa \in \Omega_f} \text{Re}(\dot{m}_{gb,i}^{(\kappa)} e^{j2\pi\kappa\tau/(N_{ht}+N_{dt})}) = m_{gb,i}^{(\tau)} & \forall i \in \Omega_{gb} \\ \sum_{\kappa \in \Omega_f} \text{Re}(\dot{h}_{chp,i}^{(\kappa)} e^{j2\pi\kappa\tau/(N_{ht}+N_{dt})}) = h_{chp,i}^{(\tau)} & \forall i \in \Omega_{chp} \\ \sum_{\kappa \in \Omega_f} \text{Re}(\dot{h}_{hp,i}^{(\tau)} e^{j2\pi\kappa\tau/(N_{ht}+N_{dt})}) = h_{hp,i}^{(\tau)} & \forall i \in \Omega_{hp} \end{cases} \quad (47)$$

$\forall \tau \in \Omega_{ht} \bigcup \Omega_{dt}$

$$\begin{cases} \sum_{\kappa \in \Omega_f} \text{Re}(\dot{p}_{n,i}^{(\kappa)} e^{j2\pi\kappa\tau/(N_{ht}+N_{dt})}) = p_{n,i}^{(\tau)} & \forall i \in \Omega_{g,n} \\ \sum_{\kappa \in \Omega_f} \text{Re}(\dot{T}_{n,i}^{(\kappa)} e^{j2\pi\kappa\tau/(N_{ht}+N_{dt})}) = T_{n,i}^{(\tau)} & \forall i \in \Omega_{h,n} \end{cases} \quad (48)$$

$\forall \tau \in \Omega_{dt}$

(4) Objective

The objective function of the ECM-based OEF model contains two parts, as expressed in (49): the first part $W_1$ is the IES operating cost, and the second part $W_2$ is a penalty on the high-frequency component magnitude for smoothing dispatch curves.

$$\min \ W_1 + W_2$$

$$W_1 = \sum_{\tau \in \Omega_{dt}} \begin{pmatrix} \sum_{i \in \Omega_{tpu}} u_{tpu,2,i}(P_{tpu,i}^{(\tau)})^2 + u_{tpu,1,i} P_{tpu,i}^{(\tau)} + u_{tpu,0,i} + \\ \sum_{i \in \Omega_{chp}} u_{chp,p2,i}(P_{chp,i}^{(\tau)})^2 + u_{chp,p1,i} P_{chp,i}^{(\tau)} + u_{chp,p0,i} + \\ \sum_{i \in \Omega_{chp}} u_{chp,h2,i}(h_{chp,i}^{(\tau)})^2 + u_{chp,h1,i} h_{chp,i}^{(\tau)} + u_{chp,h0,i} + \\ \sum_{i \in \Omega_{gw}} u_{gw,2,i}(m_{gw,i}^{(\tau)})^2 + u_{gw,1,i} m_{gw,i}^{(\tau)} + u_{gw,0,i} \end{pmatrix} \quad (49)$$

$$W_2 = \sum_{\kappa \in \Omega_f} \varepsilon \cdot \kappa \begin{pmatrix} \sum_{i \in \Omega_{gw}} |\dot{m}_{gw,i}^{(\kappa)}|^2 + \sum_{i \in \Omega_{ngu}} |\dot{m}_{ngu,i}^{(\kappa)}|^2 + \sum_{i \in \Omega_{gb}} |\dot{m}_{gb,i}^{(\kappa)}|^2 + \\ \sum_{i \in \Omega_{chp}} |\dot{h}_{chp,i}^{(\kappa)}|^2 + \sum_{i \in \Omega_{hp}} |\dot{h}_{hp,i}^{(\kappa)}|^2 \end{pmatrix}$$

where $\varepsilon$ is a tiny constant to ensure $W_2 << W_1$, so that the optimality of minimizing the operation cost is not influenced.

## 3.2 Variable space projection (VSP)

On the one hand, there are far more monitored variables than controllable variables in an OEF model because an energy network usually has numerous intermediate nodes with no connected devices; on the other hand, the monitored variables are categorized as implicit decision variables [38] that do not need to be optimized but are used for providing feasibility conditions, i.e., security constraints (44) as the bounds of these variables. This motivates us to project the monitored-variable space into the controllable-variable space through the concatenation of energy circuit constraints and time-frequency conversion constraints, which consequently expresses time-domain monitored variables as linear combinations of frequency-domain controllable variables and moves security constraints from the time-domain monitored quadrant to the frequency-domain controllable quadrant, as illustrated in Fig. 7. The details of this technique are derived as follows.

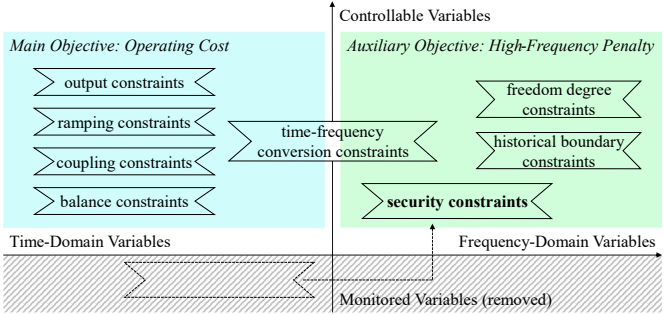

**Fig. 7** Framework of the ECM-based OEF model after VSP.

From the energy circuit constraints of the NGN in (45), we have the frequency-domain node pressure of the $\kappa$-th frequency component as expressed in (50).

$$\dot{p}_n^{(\kappa)} = (Y_{g,n}^{(\kappa)})^{-1} \dot{m}_n^{(\kappa)} \triangleq Z_{g,n}^{(\kappa)} \dot{m}_n^{(\kappa)} \tag{50}$$

where $Z_{g,n}$ is named the generalized node impedance matrix of an NGN. Note that a reference node should be removed to ensure that the generalized node admittance matrix is nonsingular, similar to the practice in power system analysis.

Then, the time-frequency conversion constraints in (48) are rewritten in matrix form, as expressed in (51).

$$\begin{bmatrix} p_{n,i}^{(0)} \\ p_{n,i}^{(1)} \\ \vdots \\ p_{n,i}^{(N_t-1)} \end{bmatrix} = \mathrm{Re}\left( \begin{bmatrix} w^{0\times 0} & w^{0\times 1} & \cdots & w^{0\times(N_f-1)} \\ w^{1\times 0} & w^{1\times 1} & \cdots & w^{1\times(N_f-1)} \\ \vdots & \vdots & \ddots & \vdots \\ w^{(N_t-1)\times 0} & w^{(N_t-1)\times 1} & \cdots & w^{(N_t-1)\times(N_f-1)} \end{bmatrix} \begin{bmatrix} \dot{p}_{n,i}^{(0)} \\ \dot{p}_{n,i}^{(1)} \\ \vdots \\ \dot{p}_{n,i}^{(N_f-1)} \end{bmatrix} \right) \tag{51}$$

where $N_t = N_{ht} + N_{dt}$ and $w = e^{j2\pi/(N_{ht}+N_{dt})}$.

Combining (50) and (51) yields the pressure of the $i$-th node in the NGN at time $\tau$:

$$p_{n,i}^{(\tau)} = \mathrm{Re}(\sum_{\kappa \in \Omega_f} w^{\tau\kappa} Z_{g,n}^{(\kappa)}[i,:]\dot{m}_n^{(\kappa)}) \tag{52}$$

Similarly, the temperature of the $i$-th node in the DHN at time $\tau$ is derived as (53).

$$T_{n,i}^{(\tau)} = \mathrm{Re}(\sum_{\kappa \in \Omega_f} w^{\tau\kappa} Z_{h,n}^{(\kappa)}[i,:]\dot{h}_n^{(\kappa)}) \tag{53}$$

where $Z_{h,n}$ is defined in (31).

Substituting (52) and (53) into the security constraints (44) yields their explicit forms:

$$\begin{cases} p_{n,i}^{lb} \leq \mathrm{Re}(\sum_{\kappa \in \Omega_f} w^{\tau\kappa} Z_{g,n}^{(\kappa)}[i,:]\dot{m}_n^{(\kappa)}) \leq p_{n,i}^{ub} & \forall i \in \Omega_{g,n}, \tau \in \Omega_{dt} \\ T_{n,i}^{lb} \leq \mathrm{Re}(\sum_{\kappa \in \Omega_f} w^{\tau\kappa} Z_{h,n}^{(\kappa)}[i,:]\dot{h}_n^{(\kappa)}) \leq T_{n,i}^{ub} & \forall i \in \Omega_{h,n}, \tau \in \Omega_{dt} \end{cases} \tag{54}$$

By using (54) to replace the original security constraints (44) in the ECM-based OEF model, which thereby removes monitored variables (i.e., $p_{n,i}^{(\tau)}$, $T_{n,i}^{(\tau)}$, $\dot{p}_{n,i}^{(\kappa)}$, and $\dot{T}_{n,i}^{(\kappa)}$), energy circuit constraints (45)-(46), and time-frequency conversion constraints for monitored variables (48), a new ECM-based OEF model is yielded as expressed in (55).

$$\begin{aligned} \min \quad & (49) \\ \text{s.t.} \quad & (37)-(43), (47), (54) \end{aligned} \tag{55}$$

It is worth noting that the reduction of variables is obtained at the cost of replacing sparse admittance matrices ($Y_{g,n}$ and $Y_{h,b}$) with dense impedance matrices ($Z_{g,n}$ and $Z_{h,n}$) in the energy circuit models, which, however, brings a sharp increase in the nonzero elements of the constraint matrix and neutralizes the efficiency improvement of variable reduction. Thus, the VSP itself does not necessarily result in better computational performance. Its real advantage is that it enables a constraint generation algorithm to exclude inactive security constraints, which significantly improves the solving efficiency.

## 3.3 Constraint generation algorithm (CGA)

After variable space projection, the security constraints in the ECM-based OEF model become dense and intractable. Regarding this issue, an important lesson from engineering practice is that not all security constraints are active for optimization: power transmission congestion is less likely to occur on non-trunk lines; under-pressure is less likely to occur on the gas-source side; and over-temperature is less likely to occur on the heat-load side. This inspires us to solve the ECM-based OEF model with security constraints added in a lazy manner, i.e., using a constraint generation algorithm. Compared with solving the OEF model directly, this algorithm is expected to remove some redundant but intractable security constraints and thereby ease the computational burden.

Specifically, all security constraints of the EPN, NGN, and DHN in (43) and (54) are first removed from (55) to obtain a raw model:

$$\begin{aligned} \min \quad & (49) \\ \text{s.t.} \quad & (37)-(42), (47) \end{aligned} \tag{56}$$

Then, this model is solved, which returns the intermediate optimal solution of node injection as $\hat{P}_n^{(\tau)}$, $\hat{m}_n^{(\kappa)}$, and $\hat{h}_n^{(\kappa)}$. Using these node injection values, the power flow on

transmission lines, pressure on NGN nodes, and temperature on DHN nodes are calculated as follows:

$$\begin{cases} \hat{p}_{n,i}^{(\tau)} = \mathrm{Re}\left(\sum_{\kappa \in \Omega_f} w^{\tau\kappa} \mathbf{Z}_{g,n}^{(\kappa)}[i,:]\hat{\dot{m}}_n^{(\kappa)}\right) & \forall i \in \Omega_{g,n}, \tau \in \Omega_{dt} \\ \hat{P}_{l,i}^{(\tau)} = \mathbf{PTDF}[i,:] \cdot \hat{\mathbf{P}}_n^{(\tau)} & \forall i \in \Omega_{e,l}, \tau \in \Omega_{dt} \\ \hat{T}_{n,i}^{(\tau)} = \mathrm{Re}\left(\sum_{\kappa \in \Omega_f} w^{\tau\kappa} \mathbf{Z}_{h,n}^{(\kappa)}[i,:]\hat{\dot{h}}_n^{(\kappa)}\right) & \forall i \in \Omega_{h,n}, \tau \in \Omega_{dt} \end{cases} \quad (57)$$

Afterward, a security check is performed to verify whether all security constraints are satisfied. The indices of the violated security constraints are recorded as expressed in (58).

$$\begin{cases} \Omega_{e,vsc}^{lb} = \{(i,\tau) \mid P_{l,i}^{(\tau)} < -P_{l,i}^{ub}, i \in \Omega_{e,l}, \tau \in \Omega_{dt}\} \\ \Omega_{e,vsc}^{ub} = \{(i,\tau) \mid P_{l,i}^{(\tau)} > P_{l,i}^{ub}, i \in \Omega_{e,l}, \tau \in \Omega_{dt}\} \\ \Omega_{g,vsc}^{lb} = \{(i,\tau) \mid p_{n,i}^{(\tau)} < p_{n,i}^{lb}, i \in \Omega_{g,n}, \tau \in \Omega_{dt}\} \\ \Omega_{g,vsc}^{ub} = \{(i,\tau) \mid p_{n,i}^{(\tau)} > p_{n,i}^{ub}, i \in \Omega_{g,n}, \tau \in \Omega_{dt}\} \\ \Omega_{h,vsc}^{lb} = \{(i,\tau) \mid T_{n,i}^{(\tau)} < T_{n,i}^{lb}, i \in \Omega_{h,n}, \tau \in \Omega_{dt}\} \\ \Omega_{h,vsc}^{ub} = \{(i,\tau) \mid T_{n,i}^{(\tau)} > T_{n,i}^{ub}, i \in \Omega_{h,n}, \tau \in \Omega_{dt}\} \end{cases} \quad (58)$$

If the above six sets are all empty, then the current model is equivalent to the original model (55), but with fewer constraints, and the current optimal solution is exactly the target solution; otherwise, the addition of more security constraints to the model should be considered for eliminating violations. Note that it is unnecessary to add all violated security constraints to the model, since severe violations can override those which are milder. In this work, a violation filtering technique that reserves the most severe $N_r$ violations for each violation type at each time point is adopted to recognize the critical security constraints. Using a superscript "*" to represent the filtered violation sets, the following security constraints in (59) are added to the model. The above procedure is repeated until no violation is detected by the security check.

$$\begin{cases} \mathbf{PTDF}[i,:] \cdot \mathbf{P}_n^{(\tau)} \geq -P_{l,i}^{ub} & \forall (i,\tau) \in \Omega_{e,vsc}^{lb*} \\ \mathbf{PTDF}[i,:] \cdot \mathbf{P}_n^{(\tau)} \leq P_{l,i}^{ub} & \forall (i,\tau) \in \Omega_{e,vsc}^{ub*} \\ \mathrm{Re}\left(\sum_{\kappa \in \Omega_f} w^{\tau\kappa} \mathbf{Z}_{g,n}^{(\kappa)}[i,:]\dot{m}_n^{(\kappa)}\right) \geq p_{n,i}^{lb} & \forall (i,\tau) \in \Omega_{g,vsc}^{lb*} \\ \mathrm{Re}\left(\sum_{\kappa \in \Omega_f} w^{\tau\kappa} \mathbf{Z}_{g,n}^{(\kappa)}[i,:]\dot{m}_n^{(\kappa)}\right) \leq p_{n,i}^{ub} & \forall (i,\tau) \in \Omega_{g,vsc}^{ub*} \\ \mathrm{Re}\left(\sum_{\kappa \in \Omega_f} w^{\tau\kappa} \mathbf{Z}_{h,n}^{(\kappa)}[i,:]\dot{h}_n^{(\kappa)}\right) \geq T_{n,i}^{lb} & \forall (i,\tau) \in \Omega_{h,vsc}^{lb*} \\ \mathrm{Re}\left(\sum_{\kappa \in \Omega_f} w^{\tau\kappa} \mathbf{Z}_{h,n}^{(\kappa)}[i,:]\dot{h}_n^{(\kappa)}\right) \leq T_{n,i}^{ub} & \forall (i,\tau) \in \Omega_{h,vsc}^{ub*} \end{cases} \quad (59)$$

The pseudocode for the whole iterative procedure is summarized in Table 3.

**Table 3** Pseudocode for constraint generation algorithm solving the ECM-based OEF model after VSP

| | |
|---|---|
| 1. | **def** ConstraintGenerationAlgorithm(model) |
| 2. | initialize model without security constraints // (56) |
| 3. | **while** True |
| 4. | $\hat{\mathbf{P}}_n^{(\tau)}$, $\hat{\dot{m}}_n^{(\kappa)}$, $\hat{\dot{h}}_n^{(\kappa)}$ ← optimize model |
| 5. | $\hat{p}_{n,i}^{(\tau)}$, $\hat{P}_{l,i}^{(\tau)}$, $\hat{T}_{n,i}^{(\tau)}$ ← energy flow calculation // (57) |
| 6. | $\Omega_{vsc}$ ← security check // (58) |
| 7. | **if** $\Omega_{vsc} = \varnothing$ // secure solution |
| 8. | **return** model, current optimal solution |
| 9. | **end if** |
| 10. | **for** violation_type in {power_ub, gas_ub, heat_ub, ...} |
| 11. | **for** $\tau$ in $\Omega_{dt}$ |
| 12. | $\Omega'_{vsc}$ ← $\Omega_{vsc}$(violation_type, $\tau$) |
| 13. | $\Omega'^*_{vsc}$ ← sort($\Omega'_{vsc}$, key=severeness)[:$N_r$] // filtering |
| 14. | **for** constraint in $\Omega'^*_{vsc}$ |
| 15. | add constraint to model // (59) |
| 16. | **end for** |
| 17. | **end for** |
| 18. | **end for** |
| 19. | **end while** |
| 20. | **end def** |

## 4. Numerical Tests

In this section, numerical tests are first performed on a small-scale IES for illustrative purposes and then performed on a large-scale IES for performance testing. In these numerical tests, the proposed ECM-based OEF model and the conventional FDM-based OEF model are compared in terms of model complexity, optimization results, and solving efficiency.

All tests are performed on a laptop with a 4-core-8-thread CPU that runs at 1.8 GHz with 8 GB of memory. Programs are coded using Python 3.8 and Gurobi 9.1.2 [39], which will be released with an open-source license at GitHub [40].

*4.1 Illustrative case: a small-scale IES*

Day-ahead dispatch with an interval of 15 minutes is solved for the IES shown in Fig. 8, which consists of a 9-bus EPN, a 7-node NGN, and a 12-node DHN. In the EPN, there are 3 power loads supplied by a CHP unit, an NGU, and a wind turbine. In the NGN, a gas load, a gas-fired boiler, and the NGU are supplied by 2 gas wells. In the DHN, there are 2 heat loads supplied by the CHP unit and gas-fired boiler. The CHP unit, NGU, and gas boiler couple the 3 energy networks. The involved data, including wind and load profiles, device parameters, and network parameters, are provided as an Excel file in [40].

To enable the ECM, a historical boundary condition of 24 hours, which is identical to that in the dispatch day, is utilized to surrogate the initial condition. To involve the FDM, spatial step lengths of 200 m for the NGN and 100 m for the DHN are adopted based on works in [27] and [29]. Under these settings, the ECM and the FDM give nearly identical results for the energy flow calculation. In addition, another hyperparameter $N_r$ in the constraint generation algorithm is set as 1 after tuning.

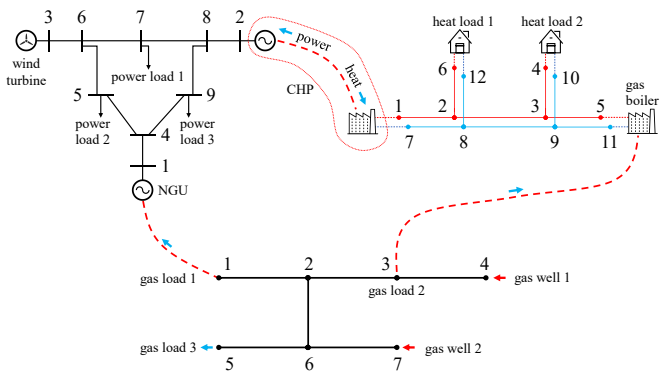

**Fig. 8** Topology of the small-scale test IES that consists of a 9-bus EPN, a 7-node NGN, and a 12-node DHN.

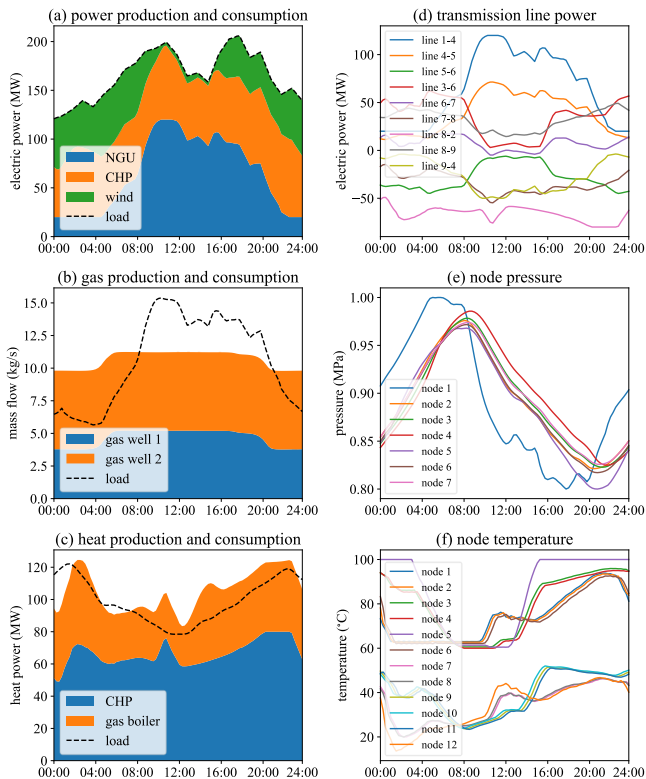

**Fig. 9** Optimization results of the ECM-based OEF model.

The optimal dispatch schedules obtained from the ECM-based OEF model, as well as the corresponding states of the monitored variables, are presented in Fig. 9. From this figure, we find that the supplies and demands in the NGN and DHN do not maintain balance in real time to realize a more flexible and efficient energy supply. Although such an imbalance results in pressure and temperature fluctuations, they are both maintained within a secure range, benefitting from the accurate modeling of the dynamic processes in the NGN and the DHN. These dispatch schedules are double-verified by the FDM-based OEF model.

Furthermore, different OEF models, which are abbreviated as the FDM model, the ECM model, the ECM+VSP model, and the ECM+VSP+CGA model in the following, are compared in terms of model complexity and computational performance.

Regarding model complexity, the constraint matrix is introduced for metrics, in which each column represents a variable, each row represents a constraint, and the entry at the $i$-th row and $j$-th column is the coefficient of the $j$-th variable in the $i$-th constraint. Note that in our implementation using the Gurobi modeling language, the constraints that contain only one variable, such as the output constraints in (37), are treated as variable bounds and do not participate in this matrix. The statistical results of the constraint matrices in these models are listed in Table 4, which indicates that the numbers of variables and constraints in the three ECM-based models are commonly reduced by an order of magnitude compared with those in the FDM model. This suggests the efficiency and simplicity of frequency-domain modeling. However, due to the denseness of energy circuit constraints and time-frequency conversion constraints, their nonzero element numbers do not show corresponding reductions. Particularly, there are more nonzero elements in the constraint matrix of the ECM+VSP model.

**Table 4** Statistical results of model complexity for the small-scale test system

|  | FDM | ECM | ECM+VSP | ECM+VSP+CGA |
|---|---|---|---|---|
| Number of Variables | 174,336 | 8,118 | 2,029 | 2,029 |
| Number of Constraints | 176,548 | 10,306 | 8,248 | 3,129 |
| Number of Nonzero Elements | 655,162 | 607,674 | 2,293,592 | 494,443 |
| Density of Constraint Matrix | 0.0021% | 0.73% | 13.71% | 7.79% |

For a more detailed comparison between the three ECM-based models, the nonzero element distributions in their constraint matrices are visualized with the same scale in Fig. 10, in which each nonzero element corresponds to a blue dot. The constraint matrix of the ECM model in Fig. 10(a) indicates that this model contains 2,029 controllable variables and 6,089 monitored variables, which verifies the statement "*there are far more monitored variables than controllable variables*" in Section 3.2. The controllable variables independently participate in 2,488 constraints of ramping, coupling, balance, and time-frequency conversion (block 1, which is common in the other two ECM-based models) and 1,728 EPN security constraints (block 3), while the monitored variables independently participate in 2,016 time-frequency conversion constraints (block 2), and they together participate in 1,358 NGN energy circuit constraints (block 4) and 2,716 DHN energy circuit constraints (block 5). From Fig. 10(a) to Fig. 10(b), the constraint matrix is visually narrowed because the ECM+VSP model removes all monitored variables. Correspondingly, the time-frequency conversion constraints for the monitored variables in block 2 are also removed, and the energy circuit constraints in block 4 and block 5 are converted into 1,344 NGN security constraints (block 6) and 1,344 DHN security constraints (block 7), which express monitored variables as linear combinations of controllable variables. As shown, the $Z$-based constraints in block 6 and block 7 are significantly denser than the *Y*-based constraints in block 4 and block 5, which accounts well for the sharp increase in the nonzero elements of the ECM+VSP model. From Fig. 10(b) to Fig. 10(c), the constraint matrix is visually shortened because the constraint generation algorithm removes most of the inactive security constraints. Specifically, there are 641 constraints in block 8 reserved from the 4,416 original security constraints, including 46 EPN security constraints from block 3 (accounting for 2.66% of the total), 128 NGN security constraints from block 6 (accounting for 9.52% of the total), and 467 DHN security constraints from block 7 (accounting for 34.74% of the total).

time and optimizing time, which benefits from a reduction in the size of the optimization model. In particular, the ECM+VSP+CGA model spends only 14.16% of the total solving time of the FDM model in this case. Another phenomenon worth noting is that the increase of nonzero elements in the model across orders of magnitude has a strong negative influence on model solvability, reflected in the ECM+VSP model having worse solving efficiency than the ECM model, although it has a constraint matrix with both smaller width and height.

**Table 5** Statistical results of model computational performance for the small-scale test system

|  |  | FDM | ECM | ECM+VSP | ECM+VSP+CGA |
|---|---|---|---|---|---|
| Objective Value ($) | | 370,996 | 370,607 | 370,607 | 370,607 |
| Time (s) | Modeling | 12.10 | 3.92 | 8.32 | 2.46 |
|  | Optimizing | 25.61 | 8.61 | 10.42 | 2.86* |
|  | Security Check | 0 | 0 | 0 | 0.02 |
|  | Total | 37.71 | 12.53 | 18.74 | 5.34 |

\* The ECM+VSP+CGA model is solved for 3 times during iteratively adding security constraints, whose computational durations are 0.76 s, 0.84 s, and 1.26 s, respectively. The total optimizing time is 2.86 s.

### 4.2 Simulated real-world case: a large-scale IES

To further verify the proposed ECM-based OEF models, day-ahead dispatch for another large-scale IES consisting of a 118-bus EPN, a 150-node NGN, and a 376-node DHN is solved. The 118-bus EPN is modified from the IEEE standard test case, which contains 186 transmission lines, 86 loads, 21 TPUs, 22 wind turbines, 6 NGUs, and 3 CHP units. The 150-node NGN is modified from an actual city-level NGN, which contains 149 gas pipelines, 51 loads, and 7 gas sources. The 376-node DHN is modified from an actual city-level DHN, which contains 515 heat pipelines, 146 heat loads, 3 CHP units, 3 gas-fired boilers, and a heat pump. The parameters of these three energy networks are scaled for balanced integration, and the same hyperparameters as those in the illustrative case are adopted in this test. The four OEF models based on the FDM and ECM all formulate reasonable optimal dispatch schedules, whose details are reported as follows.

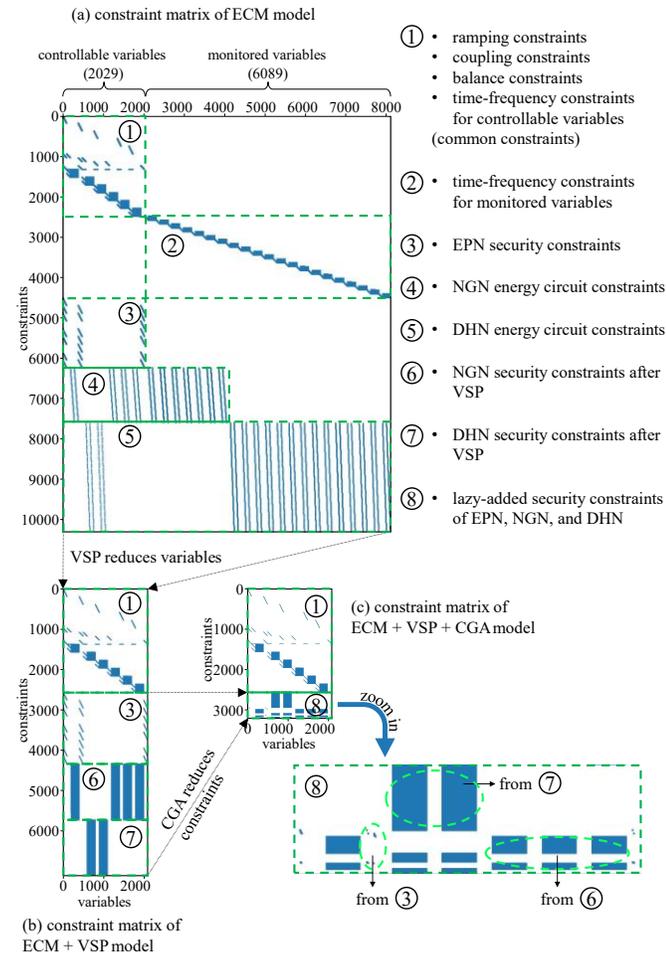

**Fig. 10** Nonzero element distributions in the constraint matrices of the three ECM-based OEF models. Note that the constraint matrix of the FDM model is too large to present here.

Regarding the computational performance, the objective values and computational time of the four OEF models are listed in Table 5. As shown, the three ECM-based models give identical optimal objective values, and the FDM model gives a similar optimal objective value with a gap less than 0.1%, which is caused by the inconsistency in the final states of the two models. Since this gap is ignorable, it is considered that the FDM model and the ECM-based models achieve the same optimality. In terms of computational time, however, the ECM-based models show a huge advantage in both modeling

Regarding the model complexity, the information about the constraint matrices of the four OEF models is listed in Table 6. Similar to the results in the illustrative case, using the FDM as a baseline, the ECM creates approximately 1/10 of the variables and constraints to model the dynamic processes in the NGN and the DHN. Then, through VSP, there are 192,849 monitored variables that occupy 94.25% of the total variables removed from the ECM model. Finally, through the CGA, there are 98,163 redundant security constraints that occupy 85.99% of the total constraints removed from the ECM+VSP model. Specifically, there are 450 EPN security constraints that account for 1.26% of the 35,712 total EPN security constraints, 288 NGN security constraints that account for 1.00% of the 28,800 total NGN security constraints, and 1,707 DHN security constraints that account for 3.46% of the 49,440 total DHN security constraints reserved in the ECM+VSP+CGA model.

**Table 6** Statistical results of model complexity for the large-scale test system

|  | FDM | ECM | ECM+VSP | ECM+VSP+CGA |
|---|---|---|---|---|
| Number of Variables | 1,960,320 | 204,608 | 11,759 | 11,759 |
| Number of Constraints | 1,998,248 | 242,114 | 114,160 | 15,997 |
| Number of Nonzero Elements | 8,987,504 | 14,794,028 | 56,074,452 | 1,592,826 |
| Density of Constraint Matrix | 0.00023% | 0.030% | 4.17% | 0.85% |

Regarding the computational performance, Table 7 records the optimal objective values and computational time of the four OEF models. It is again confirmed that the ECM-based OEF models have the same optimality as the FDM-based OEF model, since the gap between their optimal objective values is less than 0.1‰. Moreover, improved solving efficiency as a consequence of the reduced model sizes is also observed. By integrating the multiple techniques proposed in this paper, the ECM+VSP+CGA model ultimately achieves solving acceleration more than 20 times compared with the FDM model.

**Table 7** Statistical results of model computational performance for the large-scale test system

|  |  | FDM | ECM | ECM+VSP | ECM+VSP+CGA |
|---|---|---|---|---|---|
| Objective Value ($10^3$ \$) | | 15,978 | 15,991 | | 15,991 |
| Time (s) | Modeling | 165.70 | 119.18 | Numerical Trouble* | 23.66 |
| | Optimizing | 2015.88 | 412.41 | | 78.13# |
| | Security Check | 0 | 0 | | 0.80 |
| | Total | 2181.58 | 531.59 | | 102.59 |

\* Numerical trouble encountered during Gurobi solving. The solver returns an intermediate suboptimal solution. Therefore, the related data are dropped due to lack of comparability.

\# The ECM+VSP+CGA model is solved for 7 times during iteratively adding security constraints, whose computational durations are 2.91 s, 3.28 s, 10.85 s, 13.27 s, 12.78 s, 15.23 s, and 19.71 s, respectively. The total optimizing time is 78.13 s.

*4.3 Scalability test on larger-scale IESs*

To validate the scalability of the proposed OEF model and solving techniques, tests on the cascades of the large-scale IESs mentioned in the last subsection are performed. In these cascades every two elementary IESs are connected by a tie line between their EPNs and a tie pipe between their NGNs. The final variables, constraint numbers, and total solving time of the OEF models on the cascades of 1-5 elementary IESs are visualized in Fig. 11. It is observed a strictly linear growth in the numbers of variables and constraints and a slightly nonlinear growth in the total solving time as the scale expands. The nonlinearity of the latter comes from operations with polynomial time complexity, such as matrix multiplication and the interior point method. Considering these OEF problems are all solved in a quarter hour, which is far less than the call period of OEF solving, the computational time and scalability of the proposed method are acceptable.

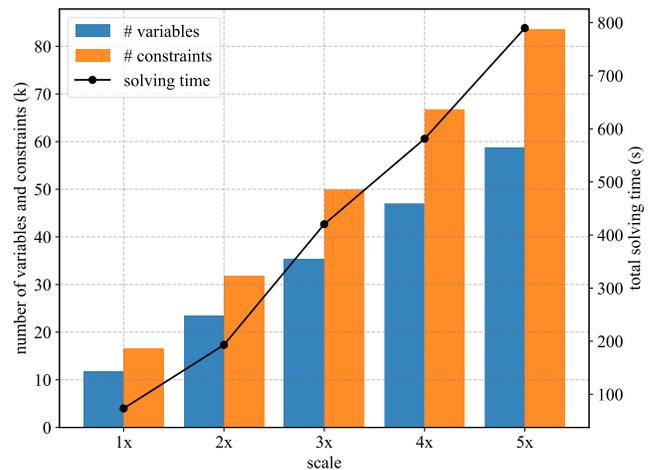

**Fig. 11** Scalability of the proposed method. The scale "1x" means an IES with a 118-bus EPN, a 150-node NGN, and a 376-node DHN; "2x" means a cascade of two such IESs; and so on. Due to the memory demand, these test results are obtained on another laptop with 16 GB of memory.

## 5. Conclusions and Future Work

This paper focuses on formulating an efficient OEF model. To this end, an energy circuit method, which models the dynamics in NGNs and DHNs in the frequency domain and algebraizes their original PDE models more efficiently than differencing in the time domain, is presented. Formulating the OEF model using this energy circuit method yields fewer variables and constraints compared with the conventional FDM-based OEF model. Furthermore, the implicit variables and redundant security constraints in this ECM-based OEF model are removed by variable space projection and the constraint generation algorithm, respectively. Through these two techniques, a more compact OEF model is obtained. Numerical tests indicate that the final ECM-based OEF model outperforms extant OEF models in terms of solving efficiency.

Considering that the IES is a promising technical area, we believe this novel OEF model will play a role in promoting IES development. To accelerate this procedure, we are willing to share our implemented works in the form of data and codes with the community.

In the future, efforts on two aspects will be devoted to deepening this work. First, the current work relies on a warm start that provides a reasonable base value and historical boundary conditions. A cold-start technique that adaptively adjusts these values will be explored. Second, more challenging OEF problems, such as security constrained OEF that considers *N*-1 contingencies, will be addressed using this efficient model.

**Acknowledgment**

This work is supported in part by the National Key Research and Development Program (2020YFE0200400), and in part by